\documentclass{iopart}

\expandafter\let\csname equation*\endcsname\relax
\expandafter\let\csname endequation*\endcsname\relax

\usepackage{lipsum} 
\usepackage{color}
\usepackage{graphicx}
\usepackage{float}
\usepackage{iopams}
\usepackage[margin=10pt,font=small,format=hang]{caption}
\usepackage{amsmath}
\usepackage{acronym}
\usepackage{multirow}
\usepackage{tabu}
\usepackage{tikz}
\usetikzlibrary{arrows,shapes,trees,decorations.pathreplacing}
\usepackage{import}
\usepackage{svn-multi}
\usepackage{hyperref}
\usepackage{mathtools}
\usepackage{url}
\usepackage{subcaption}
\hypersetup{
    colorlinks=true,
    linkcolor=blue,
    filecolor=magenta,      
    urlcolor=cyan,
}

\DeclareGraphicsExtensions{%
    .pdf,.PDF,%
    .png,.PNG,%
    .jpg,.mps,.jpeg,.jbig2,.jb2,.JPG,.JPEG,.JBIG2,.JB2}

\begin{document}

\title[]{Utilizing aLIGO Glitch Classifications to Validate Gravitational-Wave Candidates}
\author{Derek Davis$^{1,2}$,
        Laurel V. White$^1$,
        and Peter R. Saulson$^1$.
        }

\address{$^1$Syracuse University, Syracuse, NY 13244, USA}
\address{$^2$California Institute of Technology, Pasadena, CA 91125, USA}
\date{\today}

\begin{abstract}
Advanced LIGO data contains numerous noise transients, or ``glitches'', that have been shown to
reduce the sensitivity of matched filter searches for gravitational waves from compact binaries by increasing 
the rate at which random coincidences occur.
The presence of these transients has precipitated extensive work to establish
that observed gravitational wave events are
astrophysical in nature.
We discuss the response of the PyCBC search for 
gravitational waves from stellar mass binaries to various common 
glitches that were observed during Advanced LIGO's first and second observing runs. 
We show how these transients can mimic waveforms from compact
binary coalescences and quantify the likelihood that a given class
of glitches will create a trigger in the search pipeline. 
We explore the specific waveform parameters that are most similar 
to different glitch classes
and demonstrate how knowledge of these similarities
can be used when evaluating the significance
of gravitational-wave candidates.
\end{abstract}

\maketitle

\section{Introduction}\label{s:intro} 

Advanced LIGO's (aLIGO) first and second observing runs (O1 and O2) have allowed
gravitational waves to be detected for the first time.
During these obsevring runs, aLIGO observed gravitational-wave 
signals from at least 10 binary black hole (BBH) systems \cite{gwtc-1} and 1
binary neutron star (BNS) system \cite{GW170817}. 

The detection of the first BNS, GW170817, was initially complicated by the presence of a loud instrumental
noise transient that had to be removed before analysis could be completed
\cite{GW170817,Pankow:2018mit}. While a somewhat unlikely coincidence, this event highlighted the nature
of aLIGO data as neither Gaussian nor stationary over long time periods 
\cite{walker:2018,Nuttall:2015dqa,GW150914_detchar}.
The presence of noise transients, generally referred to as ``glitches'', have been shown to
impact the sensitivity of searches for gravitational waves from 
compact binary coalescence (CBC) by 
mimicking the appearance of a gravitational wave to matched filter 
based searches \cite{GW150914_detchar,O1DQ,nitz:2018sg}. 
When a candidate signal is identified by a search pipeline, rigorous
studies can be undertaken regarding the time in question to 
understand if the trigger is related to instrumental causes
\cite{gwtc-1,Nuttall:2018dq,Berger:2018dq}.
Tests to quickly evaluate the data quality
around a candidate signal 
before initiating an 
additional search for an electromagnetic (EM)
counterpart were  
routinely completed as
part of the O2 EM follow up process \cite{O2mma}.

Previous work on detector characterization of gravitational-wave
interferometers \cite{gwtc-1,GW150914_detchar,GravitySpy,Powell:2015class,Powell:2017classii,Mukund:2017,Mukherjee:2010} 
has identified a number
of classes of glitches 
that are present in the data. 
These classifications have allowed valuable follow up using auxiliary sensors
\cite{Nuttall:2018dq,Berger:2018dq,Smith:2011hv,Essick:2014wwa,Ajith:2014bcv,Isogai:2010,cabero:2019bg}
to identify causes of each class and improve
the overall gravitational-wave data quality. 
If a known glitch class can be associated with a well 
understood witness sensor, mitigation of the glitch may be possible
either through instrumental intervention or by
identifying these periods of data as corrupted. 
Unfortunately, there is a number of identified classes of glitches 
for which mitigation methods are not yet understood. 
For these glitch classes, understanding how searches can separate 
instrumental transients from similar astrophysical signals
is the highest priority. 

This paper demonstrates how the most distinguishable glitch classes affect matched filter
searches for gravitational waves from CBC sources
and how to evaluate the significance of
gravitational-wave candidates near glitches.
In order to identify periods that are corrupted by
known classes of glitches, we take advantage of Gravity Spy,
a machine-learning-based image classifier \cite{GravitySpy}. 
We examine the response to these glitches by the
PyCBC search pipeline, one of the pipelines used to find CBC signals with aLIGO
\cite{Usman:2015kfa,Nitz:2017stat,pycbc-software}. 
We show how these glitches can mimic waveforms from astrophysical
sources of gravitational waves and quantify the likelihood of a given 
glitch from each glitch class
to create a significant trigger in PyCBC.
We then demonstrate how measurements of the background 
during times coincident with glitches
can be used to evaluate if candidate signals are consistent
with the expected response of the search to a population
of glitches.   

\section{PyCBC Detection Statistic}\label{sec:pycbc} 

In this work we utilize the PyCBC \cite{Usman:2015kfa,Nitz:2017stat,nitz:2018sg} 
search pipeline as an example
matched filter search for CBC signals.
While this work may be broadly applicable
to other analysis pipelines used to identify CBC
signals with matched filtering \cite{Messick:2017gstlal,venumadhav:2019ias},
differences in how pipelines rank 
candidates may 
result in slight differences in the effect of
glitches on the analysis. 
An overview of how PyCBC identifies and ranks significant 
candidate triggers follows. 

The PyCBC search pipeline is designed to identify
gravitational waves generated by compact binary coalescences
in interferometer data. 
To evaluate the significance of an individual candidate, 
each trigger is ranked based on the
PyCBC detection statistic.
While the complete PyCBC detection statistic is designed for identifying
triggers that are coincident between detectors, 
we consider primarily the single detector detection statistic.
This has three components: 
the signal-to-noise ratio (SNR), 
the chi-squared ($\chi^2$) discriminator, 
and the sine-Gaussian discriminator. 

The core of the detection statistic is the signal-to-noise ratio 
for a given template.
The SNR for a matched filter with a specific waveform template $h$ is

\begin{equation}
 \rho^2 \equiv \frac{\|\left\langle s | h \right\rangle\|^2}{\left\langle h | h \right\rangle}
 \text{ ,}
\end{equation}
\label{eq:snr}
where the inner product is given by

\begin{equation}
 \langle a|b\rangle = 4 \int^\infty_0 \frac{\tilde{a}(f)\tilde{b}^*(f)}{S_n(f)} df
 \text{ ,}
\end{equation}
with $s$ the strain data and $S_n(f)$ the measured power spectral 
density for the time in question.

If aLIGO noise was perfectly Gaussian, the matched
filter SNR alone would be the optimal 
detection statistic.
However, since the data contains non-Gaussian fluctuations \cite{GW150914_detchar,O1DQ}, 
numerous additional signal consistency tests are required to 
discriminate between instrumental artifacts
and astrophysical signals.  
In an ideal signal consistency test,
noise triggers are assigned a lower ranking statistic
than comparably loud astrophysical signals.

To provide this discriminatory power, one of the most useful tests 
for gravitational-wave signals is
the chi-squared discriminator \cite{Allen:2005fk}.
The test is constructed by dividing the frequency space
spanned by the waveform template into bins of equal power, 
and checking if each bin contributes
the expected amount of power. 
This gives a measure of how well 
a candidate trigger matches the signal morphology 
of the template. 
Specifically, the chi-squared discriminator for a trigger is given by

\begin{equation}
\chi^2_r = \frac{1}{2p-2}\sum_{i=1}^{p} \left\|\langle s|h_i \rangle - \langle h_i|h_i \rangle\right\|^2,
\end{equation}
with $h_i$ the waveform template in a given frequency bin.

This value should follow a reduced $\chi^2$ distribution with $2p-2$ degrees of freedom. 
The choice of $p$ is scaled based on the duration of each template, 
so that a sufficient number of bins with measurable power are used. 
If the value of the chi-squared test is greater than unity, 
the detection statistic for the related trigger is reduced 
to produce a ``re-weighted SNR'', $\tilde{\rho}$.
This is

\begin{equation}
 \tilde{\rho} = \begin{cases} 
        \rho & \mathrm{for}\ \chi^2_r \leq 1 \\
        \rho\left[ \frac{1}{2} \left(1 + \left(\chi^2_r\right)^3\right)\right]^{-1/6} & 
        \mathrm{for}\ \chi^2_r > 1
    \end{cases}.
\end{equation}

The effectiveness of the chi-squared discriminator has been shown to be
dependent on the duration of the signal and the number of
bins used in the test \cite{O1DQ}.
For long duration signals, the test provides excellent rejection of many classes 
of glitches.
For short duration signals, this test has reduced efficiency. 
To help address this effect for short duration templates, 
an additional signal consistency test, the sine-Gaussian discriminator,
is utilized.

The sine-Gaussian discriminator is designed to downrank triggers with 
excess power at frequencies above the expected maximal
frequency of the signal at merger \cite{nitz:2018sg}. 
If excess power is detected above this frequency, 
the trigger is not likely to be generated by a CBC signal. 
To quantify the excess power present at high 
frequencies, a number of sine-Gaussian 
wavelets with frequencies
above this maximum are matched filtered against the data.
These wavelets are parameterized by their 
frequency, $f_0$, 
central time $t_0$,
and quality factor $Q$.
In the time domain, each wavelet can be written as

\begin{equation}
    g(t) =  \exp \left(
         -4 \pi f_0^2 \frac{(t-t_0)^2}{Q^2}
         \right)
         \cos ( 2 \pi f_0 t + \phi_0 )
    \text{ .}
\end{equation}

A new signal discriminator can be written down as the 
sum of the measured matched filter
SNR squared of each individual sine-Gaussian tile.
In the case of $N$ different tiles, this is

\begin{equation}
  \chi^2_{r,sg} \equiv \frac{1}{2N} \sum_{i=1}^{N} \rho_i^2 
  = \frac{1}{2N} \sum_{i=1}^{N} \langle s| \tilde{g}_i(f,f_0,t_0,Q)\rangle^2
  \text{ .}
\end{equation}

Similar to the chi-squared discriminator, this statistic should
follow a reduced $\chi^2$ distribution with $2N$ degrees of freedom
for astrophysical signals. 
The result of this test is then used to compute a new detection statistic, 
$\tilde{\rho}_{sg}$, defined as

\begin{equation}
 \tilde{\rho}_{sg} = \begin{cases} 
        \tilde{\rho} & \mathrm{for}\ \chi^2_{r,sq} \leq 4 \\
        \tilde{\rho}\left( \chi^2_{r,sq} / 4 \right)^{-1/2} & 
        \mathrm{for}\ \chi^2_{r,sq} > 4
    \end{cases}
    \text{ .}
\end{equation}
The value of $4$ (as opposed to 1) is chosen as the 
threshold to account for the 
expected variability of $\tilde{\rho}_{sg}$ in Gaussian noise. 
Values above $4$ are indicative of likely non-Gaussian features in the data. 

Even with these consistency tests, many glitches
still produce significant triggers in the search.
Extensive investigation is done to identify  
periods corrupted by problematic glitches 
and and remove these time segments from the analysis \cite{O1DQ}.
However, these analyses are primarily done
with auxiliary witness sensors, 
and do not rely upon the gravitational-wave strain data 
to identify glitches.
One of the reasons for this current lack of 
investigations based on analyses of the strain data is the need for 
a robust classification method for glitches
that is not based on the matched filtering
pipeline itself. 

\section{Gravity Spy Classification}\label{sec:grav_spy} 

In order to classify glitches in this work, we use 
Gravity Spy \cite{GravitySpy,gravityspy-ts},  
a machine learning based classification tool that utilizes citizen science efforts. 
Gravity Spy has been used to quantify glitch rates
and identify large sets of similar glitches
\cite{Bahaadini:2017dqg,PhysRevD.99.082002,alog:gsscratchySC,alog:gsDove},
as it can quickly and
accurately identify common classes of instrumental artifacts in the detector.
These studies have generally been aimed
at understanding detector performance and the sources
of these glitch classes.
However, Gravity Spy can also be used to study 
the impact of glitches on searches for gravitational waves as it provides
a method to develop an initial dataset of glitches to investigate
that is independent of the search methodology. 
This section explains the data selection process for
the Gravity Spy pipeline
including possible selection effects relevant to this study.
Full details on the classification methods for Gravity Spy can be 
found in \cite{GravitySpy}.

\begin{figure}
\centering
    \begin{subfigure}{0.45\textwidth}
        \centering
        \includegraphics[width=\linewidth]{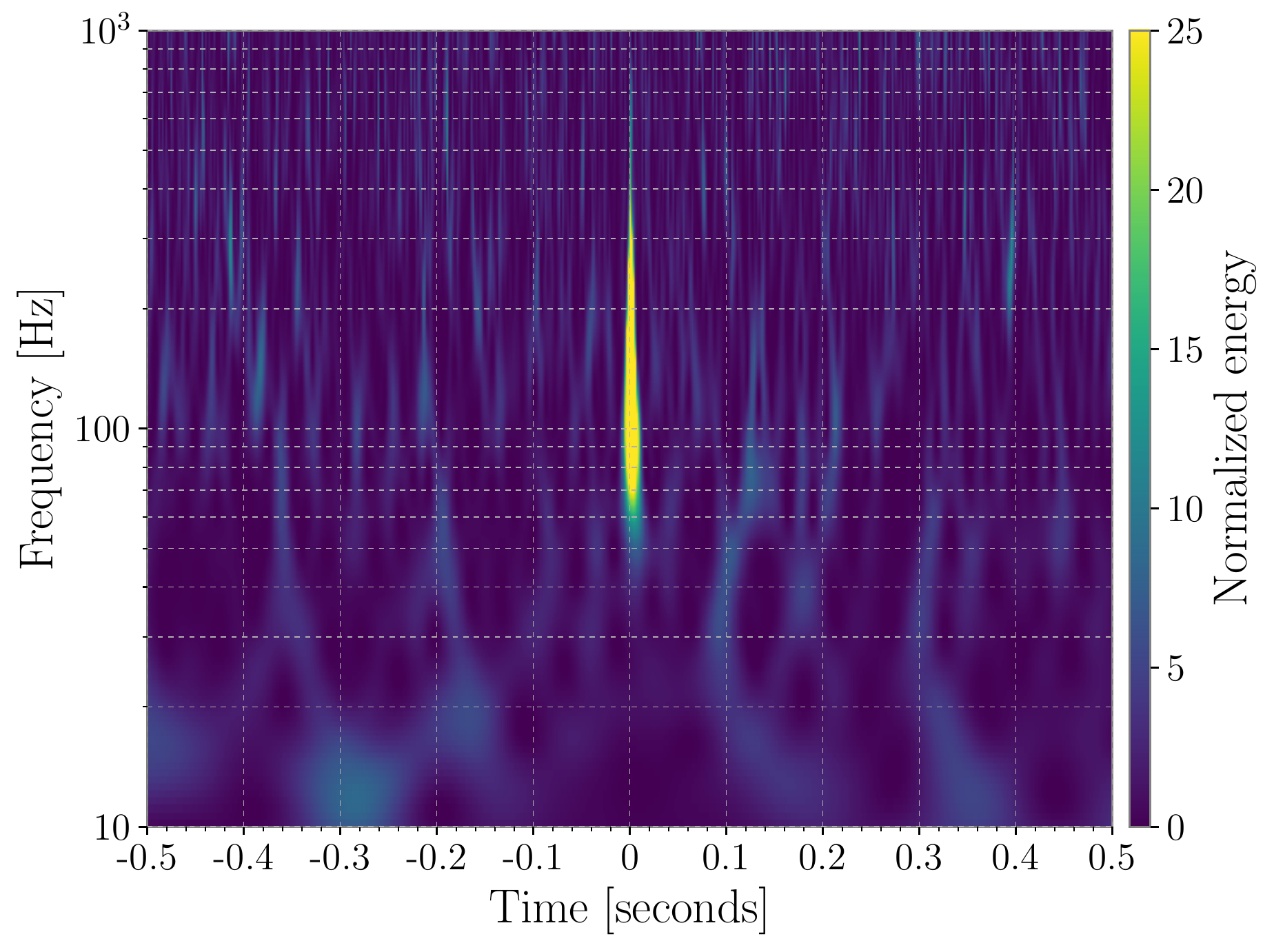}
        \caption{Blips}\label{fig:scan_a}
    \end{subfigure} %
    \begin{subfigure}{.45\textwidth}
        \centering
        \includegraphics[width=\linewidth]{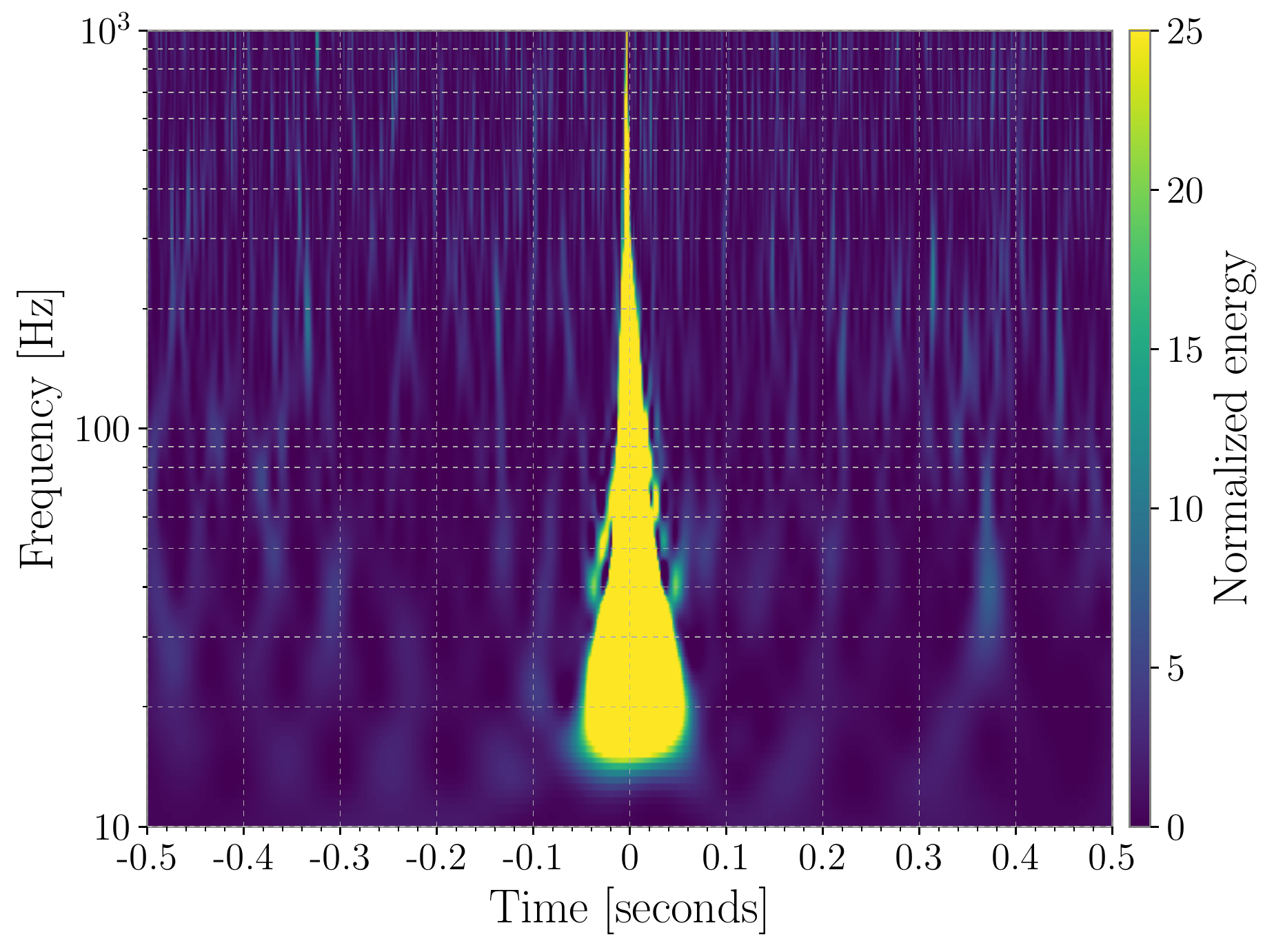}
        \caption{Koi Fish}\label{fig:scan_d}
    \end{subfigure}
    \begin{subfigure}{.45\textwidth}
        \centering
        \includegraphics[width=\linewidth]{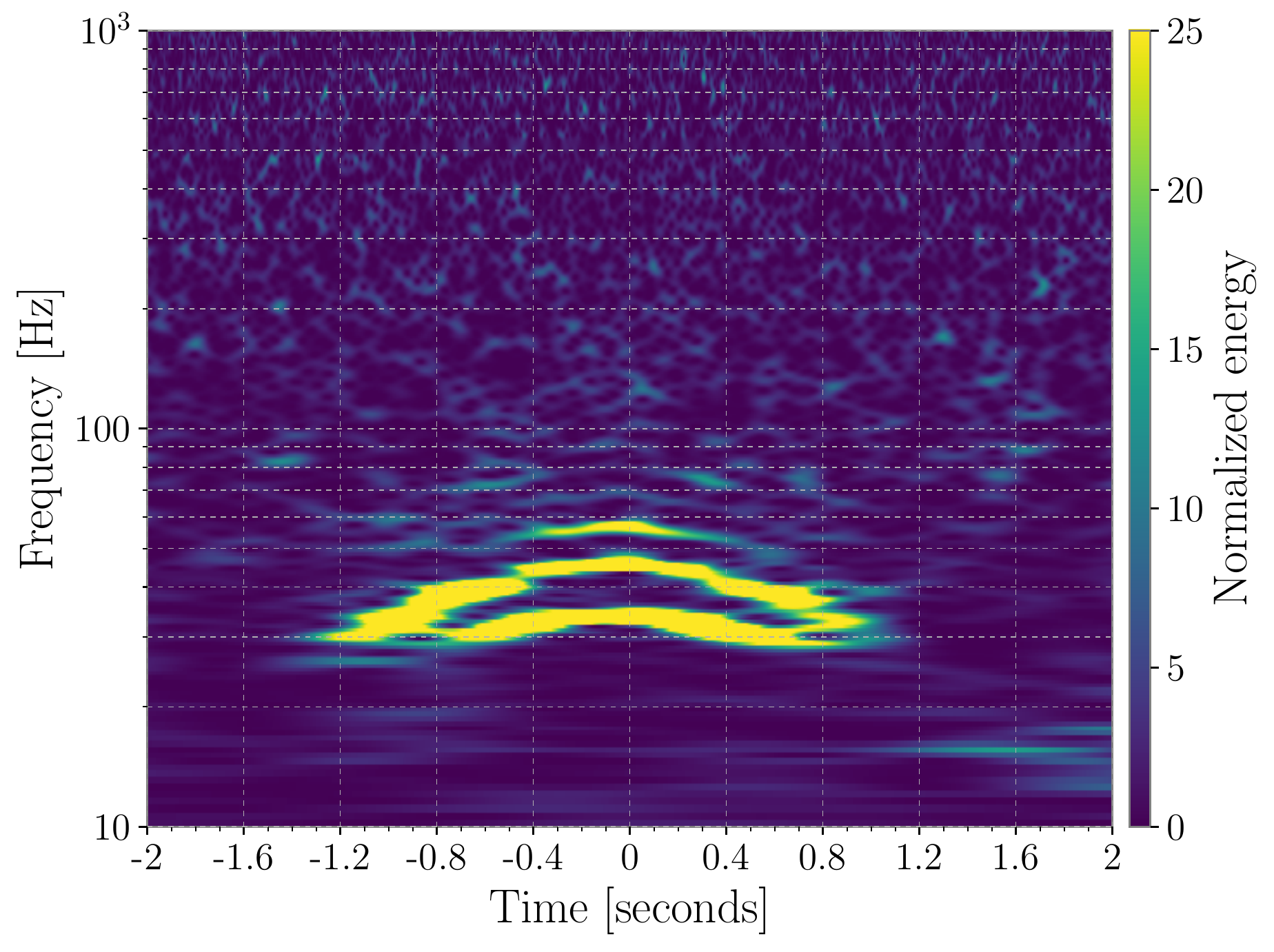}
        \caption{Scattered Light}\label{fig:scan_b}
    \end{subfigure} %
    \begin{subfigure}{.45\textwidth}
        \centering
        \includegraphics[width=\linewidth]{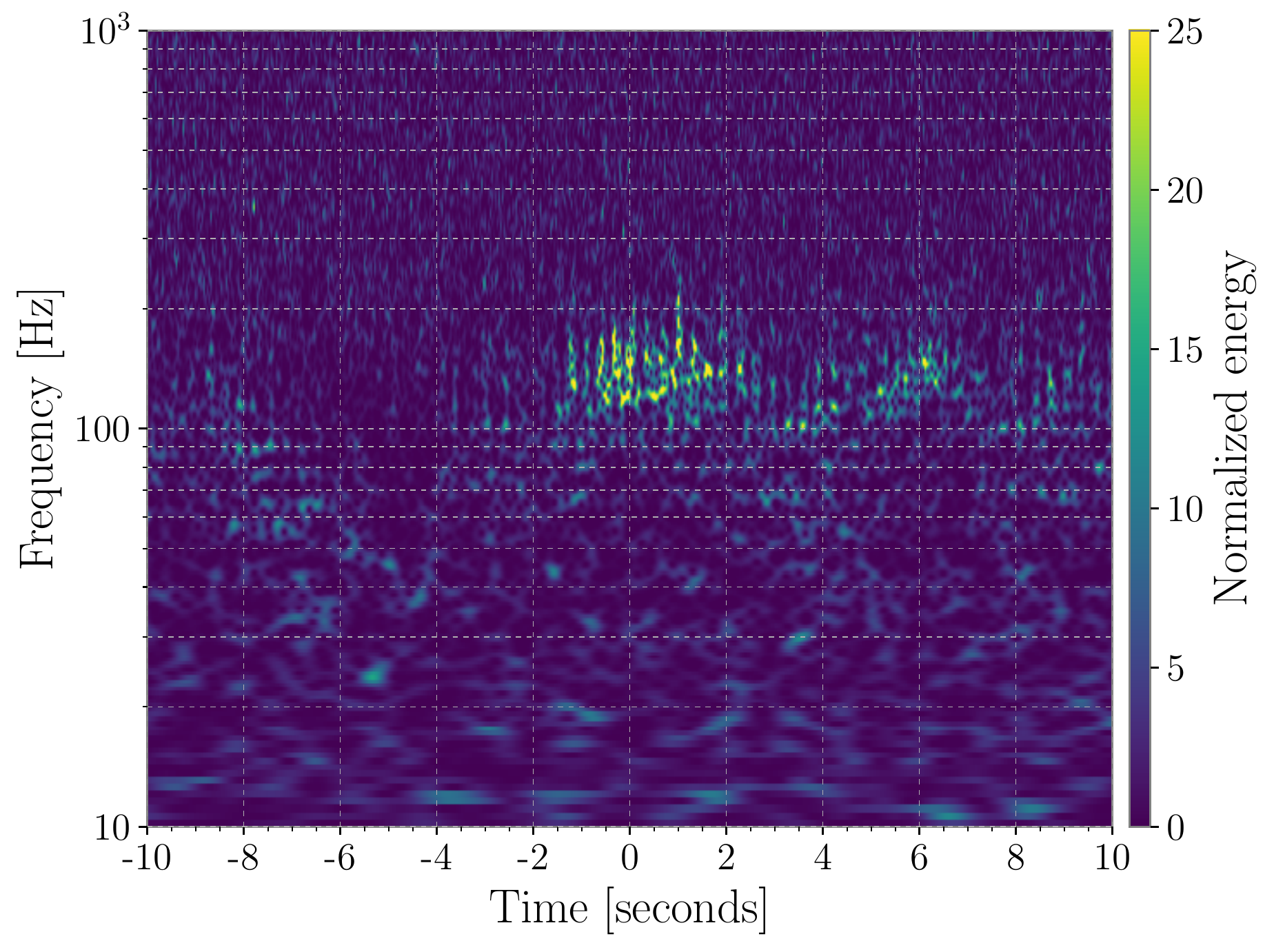}
        \caption{Scratchy}\label{fig:scan_c}
    \end{subfigure}
\caption{Spectrograms of the four
problematic Gravity Spy glitch classes discussed in this work.
The glitch classes blips (top-left), koi fish (top-right),
scattered light (bottom-left) and scratchy (bottom-right) are highlighted here due to their
known impact on the PyCBC search. 
Note the diversity in duration and morphology among these four glitch classes.
}
\label{fig:scan}
\end{figure}

At its core, Gravity Spy is an image classifier based on 
convolutional neural network methods \cite{Krizh:2012cnn,Bahaadini:2018nkh}.
Before the classifier can be applied, 
time periods containing glitches must be identified 
and the relevant detector data translated into an image format that
the neural network can process.
To identify a glitch, the Gravity Spy pipeline takes advantage of the 
Omicron pipeline \cite{Robinet:2015om}. 
Omicron uses a set of sine-Guassian wavelets to identify
excess power in detector data.
Any Omicron trigger with an SNR above 7.5 is reported to the pipeline.
Once a time window containing a glitch or astrophysical signal is identified, 
the time series data is transformed into a 
spectrogram using the Q-transform \cite{Chatterji:2004qg}.
This representation provides the input that both the machine learning classifier and 
citizen scientists will use in their classification efforts. 
This glitch image is fed into the classifier and a confidence
score ranging from $0.0$ to $1.0$ is given for each category. 
The total sum over all categories is $1.0$, with the highest
numeric value representing the most likely classification.

The SNR threshold from Omicron is used to increase the chance that a clearly
defined glitch will be visible in the spectrogram representation and
decrease the overall size of the dataset. 
In the context of using these classifications for understanding
the effect of glitches on search pipelines, this does provide some bias, 
as noise sources problematic to the searches may
not meet this threshold. 
This consideration is especially important for long duration signals
that are not expected to be identifiable in this representation,
but can be found using matched filtering. 

Another important consideration is the set of possible classifications 
that Gravity Spy can provide. 
While the list of possible classifications the pipeline can assign
includes ``None of the Above'' and ``No Glitch'' classes,
the classification is mostly limited
to predetermined classes from a training set \cite{GS_glitch_dataset,gravityspy-ts}.
Therefore, if a glitch  unknown to the pipeline is classified, 
the result has a much higher chance of being incorrect, thus 
 contaminating the glitch set. 
To guard against this problem, we set a minimum confidence
of $0.95$ for all glitch classes to reduce the risk of contamination.

Of the classes Gravity Spy has in its training set, we will focus on four 
in this work: ``blip'', ``koi fish'', ``scattered light'', and ``scratchy''.
spectrograms of representative examples of each of these glitches
are shown in Figure \ref{fig:scan}.
These four are chosen as they have been previously identified as problematic
for searches for gravitational waves from compact binaries
\cite{gwtc-1,O1DQ}.
These glitches are also  some of the most common glitches
in the LIGO detectors,
allowing for a broad statistical study. 
Finally, each of these classes has yet to be completely mitigated 
via instrumental means in the LIGO detectors. 
Due to this, it is likely these glitches will be present
in future observing runs and continue to limit the sensitivity of
searches.
Further discussion of these glitch classes can be found in 
\cite{gwtc-1,O1DQ,GW150914_detchar}.

\section{How different kinds of glitches mimic traits of signals}\label{sec:bank} 

PyCBC signal consistency tests have been shown to discriminate well
between glitches and astrophysical signals 
\cite{O1DQ,nitz:2018sg,Usman:2015kfa,Nitz:2017stat,Allen:2005fk}. 
However, the wide range of template parameters included in the 
search \cite{dalcanton:2017tb}, 
combined with the wide variety of instrumental artifacts,
means that this discriminating power is not uniformly effective 
across the entirety of the search 
parameter space \cite{O1DQ}.

\begin{figure}[h]
\centering
    \includegraphics[width=\textwidth]{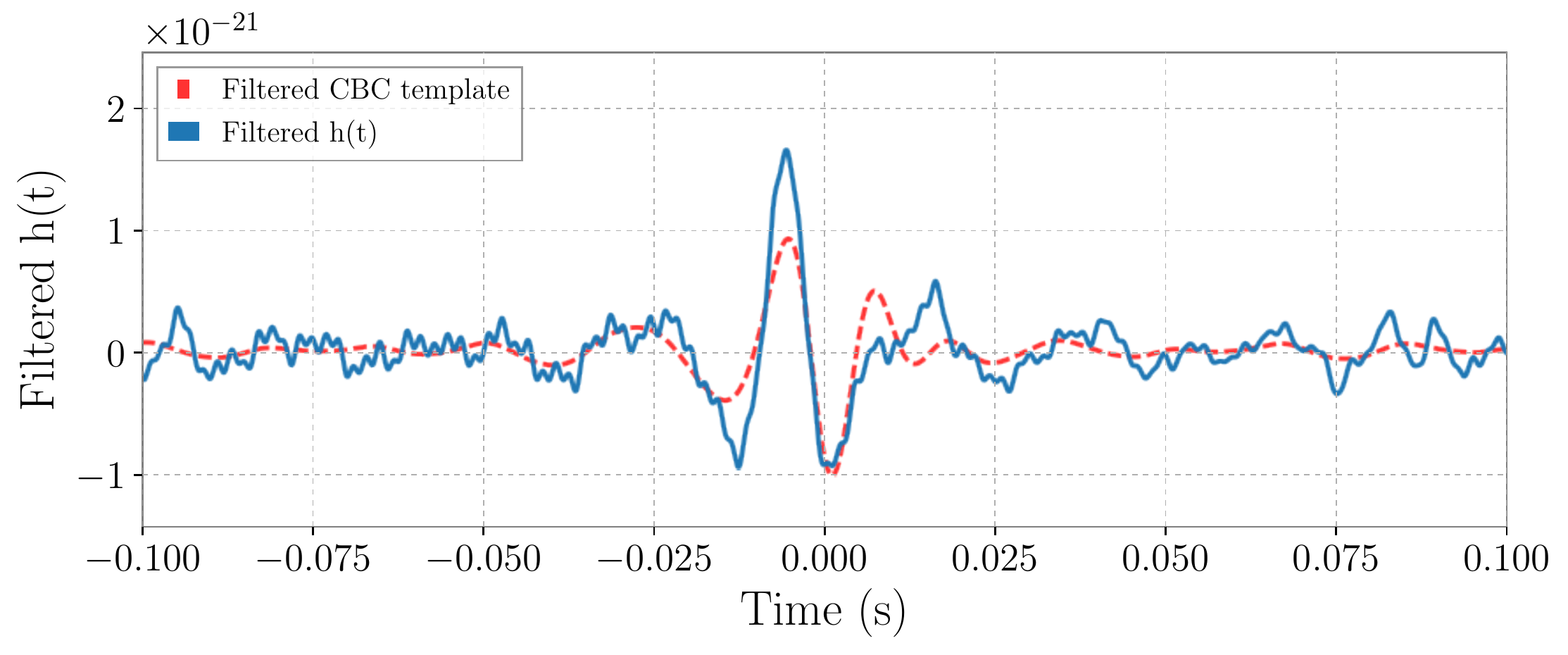}
    \caption{An overlay of a timeseries of a specifically chosen high-mass CBC template and detector data around a blip glitch. %
    Both of the timeseries have been filtered with bandpass filters to isolate the most sensitive region of 35-500 Hz and
    notch filters to repress noise from calibration lines
    and harmonics of the 60 Hz power line. %
    This visualization serves to show the similarity between a blip glitch %
    and a CBC template after the response of the detector is considered.}
    \label{fig:template}
\end{figure}

This concern is easily demonstrated 
for blip glitches.
When plotted against a timeseries of the data around
a representative blip glitch, there is significant overlap similarity between
a blip and a the best matching template in the PyCBC bank, as shown in 
Figure \ref{fig:template}. 
A blip waveform has a few short, loud cycles, similar 
to a sine-Gaussian pulse 
\cite{GW150914_detchar,nitz:2018sg,cabero:2019bg}. 
The detector is less sensitive at lower frequencies, 
meaning a template that reaches merger by 100 Hz
will be in the observable band for only a few
gravitational wave cycles and qualitatively 
match the model for a representative blip.
Such templates correspond to some of the most 
massive systems in the PyCBC template bank.
As blip glitches do not occur coincident in
both detectors, they are known to be instrumental 
artifacts. 
However, these glitches have been noted as one of the limiting
sources of noise for searches for 
gravitational waves from high mass
binary black holes \cite{O1_IMBH,GW150914_no_assump}.

To identify PyCBC triggers coincident with a glitch, 
we analyzed short segments of data 
around glitches identified by
Gravity Spy as belonging to a specific glitch class, and recorded all 
PyCBC triggers that met a minimum detection statistic value. 
Multiple seconds of data after the glitch were included to ensure that
the full duration of the glitch was included and that
triggers intersecting the inspiral component of the waveform were correctly
recorded.
This duration was different for each glitch class;
a 2 second window was used for blips and koi fish, 
4 seconds for scattered light, 
and 30 seconds for scratchy.
In cases when the time window for neighboring glitches intersected, 
this was considered one glitch. 
These single-detector PyCBC triggers that are
idenitifed coincident with Gravity Spy glitches are used throughout this 
work to explore how glitches can mimic CBC signals.
Using this set of PyCBC triggers during times identified
by Gravity Spy, we examine which CBC templates are most likely to give
a significant false positive to each kind of glitch.


\begin{figure}[t]
\centering
    \begin{subfigure}{0.45\textwidth}
        \centering
        \includegraphics[width=\linewidth]{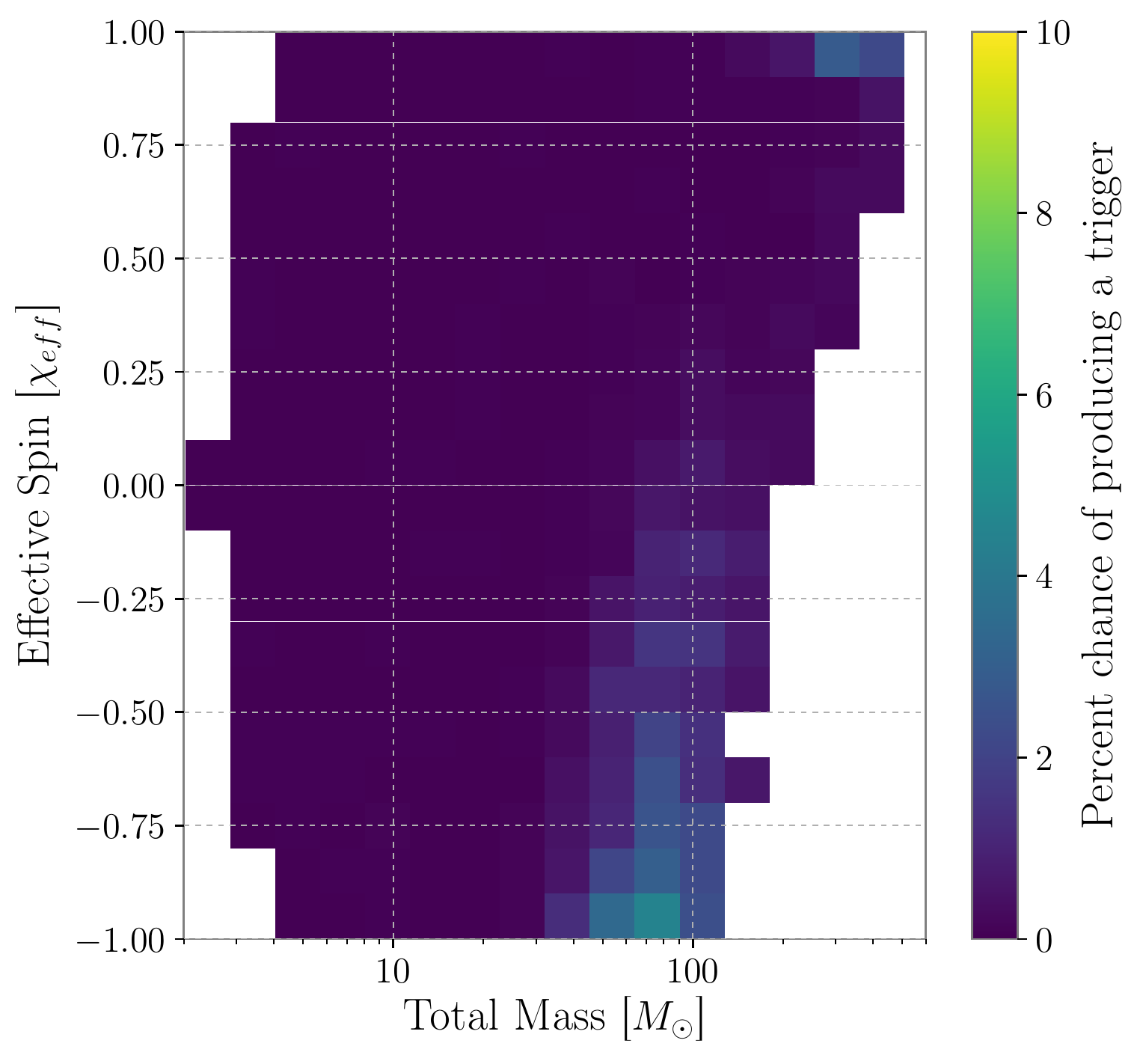}
        \caption{Blips}\label{fig:prob_MX_a}
    \end{subfigure} %
    \begin{subfigure}{.45\textwidth}
        \centering
        \includegraphics[width=\linewidth]{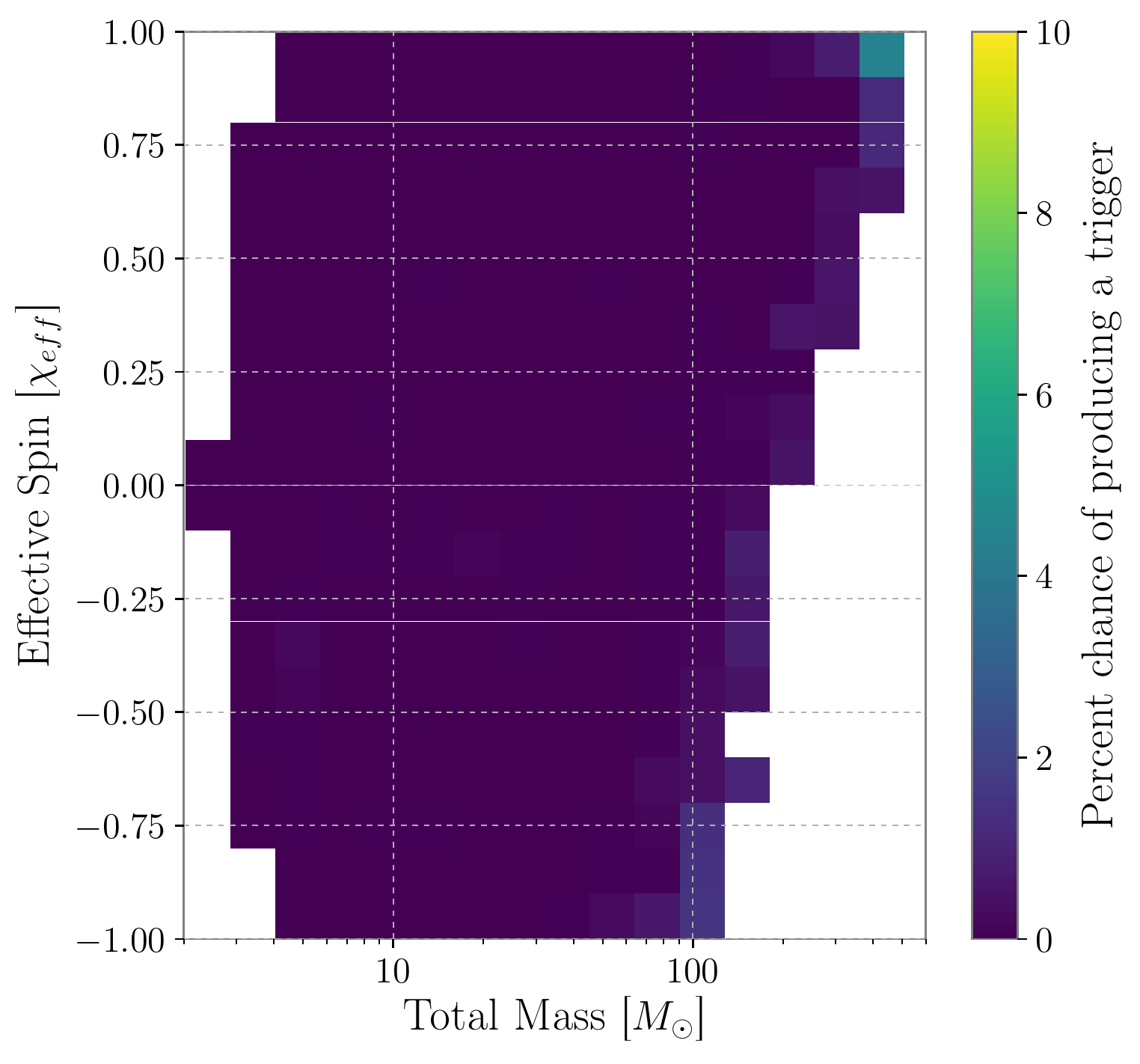}
        \caption{Koi fish}\label{fig:prob_MX_d}
    \end{subfigure}
    \begin{subfigure}{.45\textwidth}
        \centering
        \includegraphics[width=\linewidth]{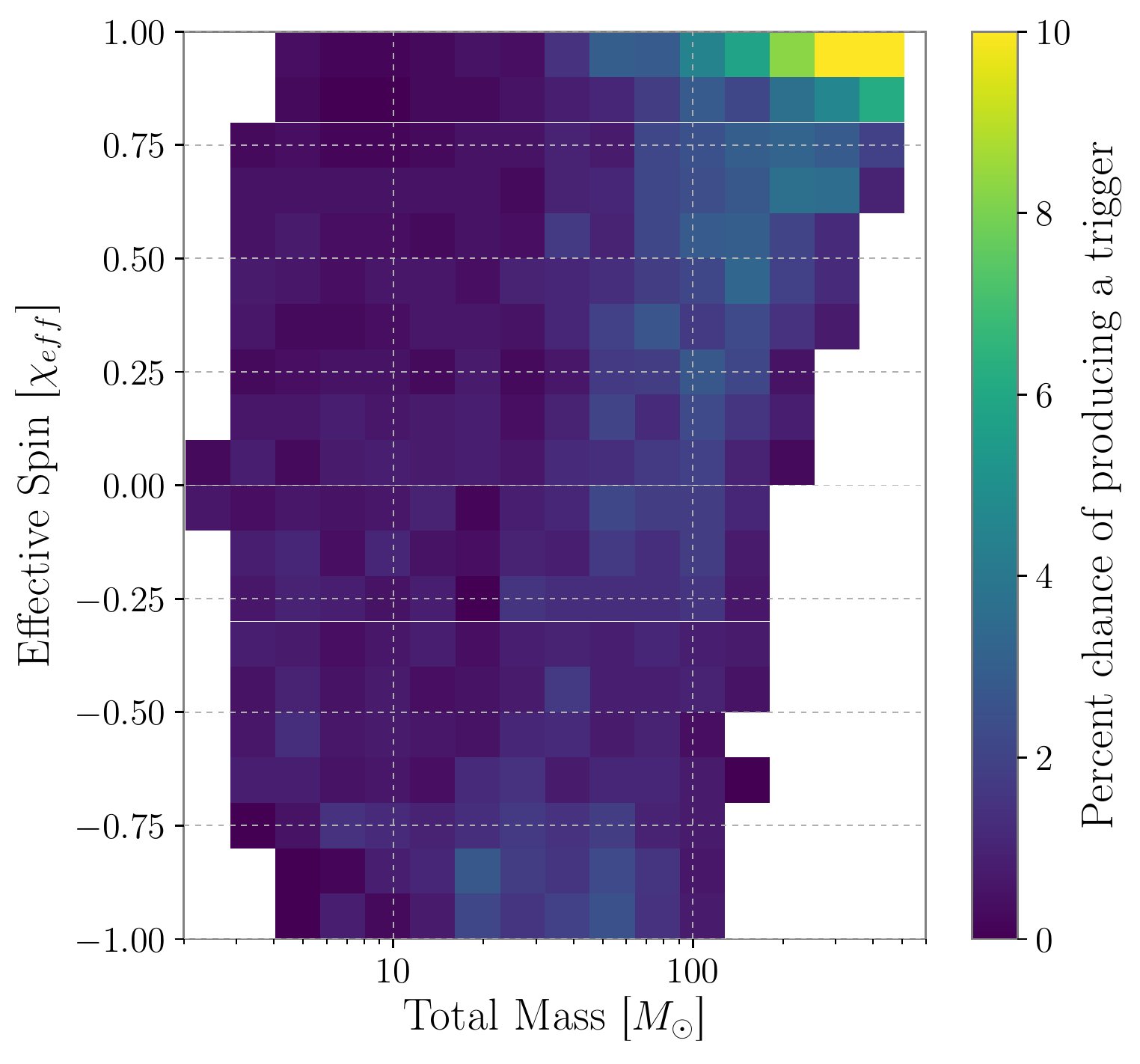}
        \caption{Scattered Light}\label{fig:prob_MX_b}
    \end{subfigure} %
    \begin{subfigure}{.45\textwidth}
        \centering
        \includegraphics[width=\linewidth]{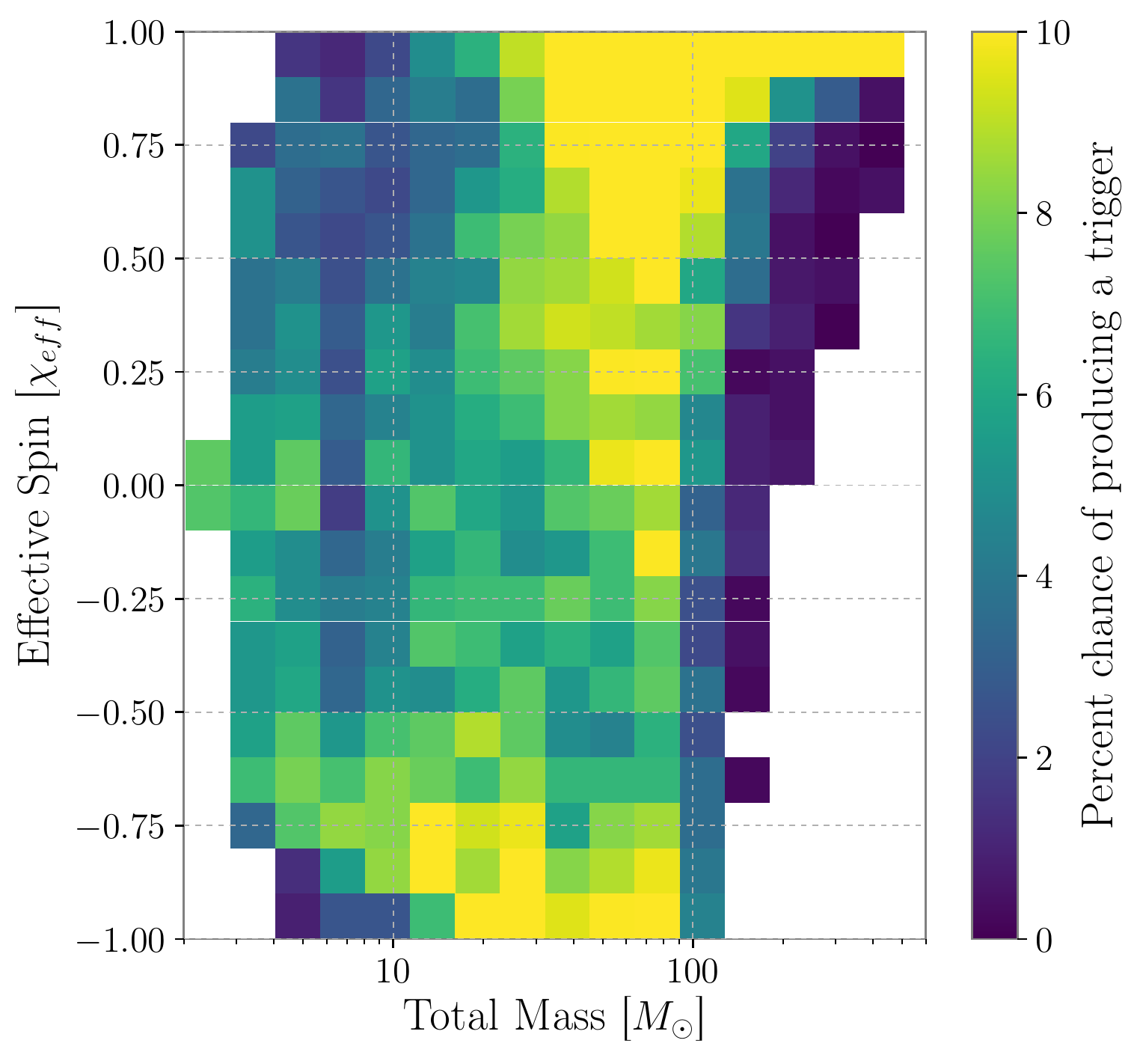}
        \caption{Scratchy}\label{fig:prob_MX_c}
    \end{subfigure}
\caption{Probability of producing a trigger above sine-Gaussian SNR of 7.0 in specific regions of the template bank parameter space for each glitch class.
Triggers are binned by total mass and effective spin of the corresponding template.
Blips (top-left), koi fish (top-right),
scattered light (bottom-left) and scratchy (bottom-right) each show maximum probabilities
in different parts of the parameter space.
} 
\label{fig:prob_MX}
\end{figure}

For each class of glitch in our data set, 
we test to see the likelihood that a single glitch from this class produces a trigger
above a detection statistic threshold (in this case 
$\tilde{\rho}_{sg}>7.0$ in each bin)
in each part of the template bank parameter space. 
This value was chosen because a trigger of this detection statistic 
value combined with the minimal value possible in the other 
detector ($\tilde{\rho}_{sg} \approx 5.5$) would result in a candidate with 
a combined network detection statistic of 
$\tilde{\rho}_{sg,net}>9.0$, strong enough to be identified separate from the background at a false alarm rate of approximately 1 per year. 
We choose to bin the parameter space spanned by the O2 PyCBC template bank
uniformly in both $\chi_{eff}$ (the effective spin of the system) and $\log (M_{total})$ to help account for the
reduced density of templates at high $M_{total}$.
The probability of a glitch producing a trigger in each bin
is highly dependent on the total number of templates
within the bin, and hence the size of the bin itself. 
However, the regions of the parameter space that are identified as the
most likely to produce a trigger are robust to choice of bin size.

\begin{figure}[t]
\centering
    \begin{subfigure}{0.45\textwidth}
        \centering
        \includegraphics[width=\linewidth]{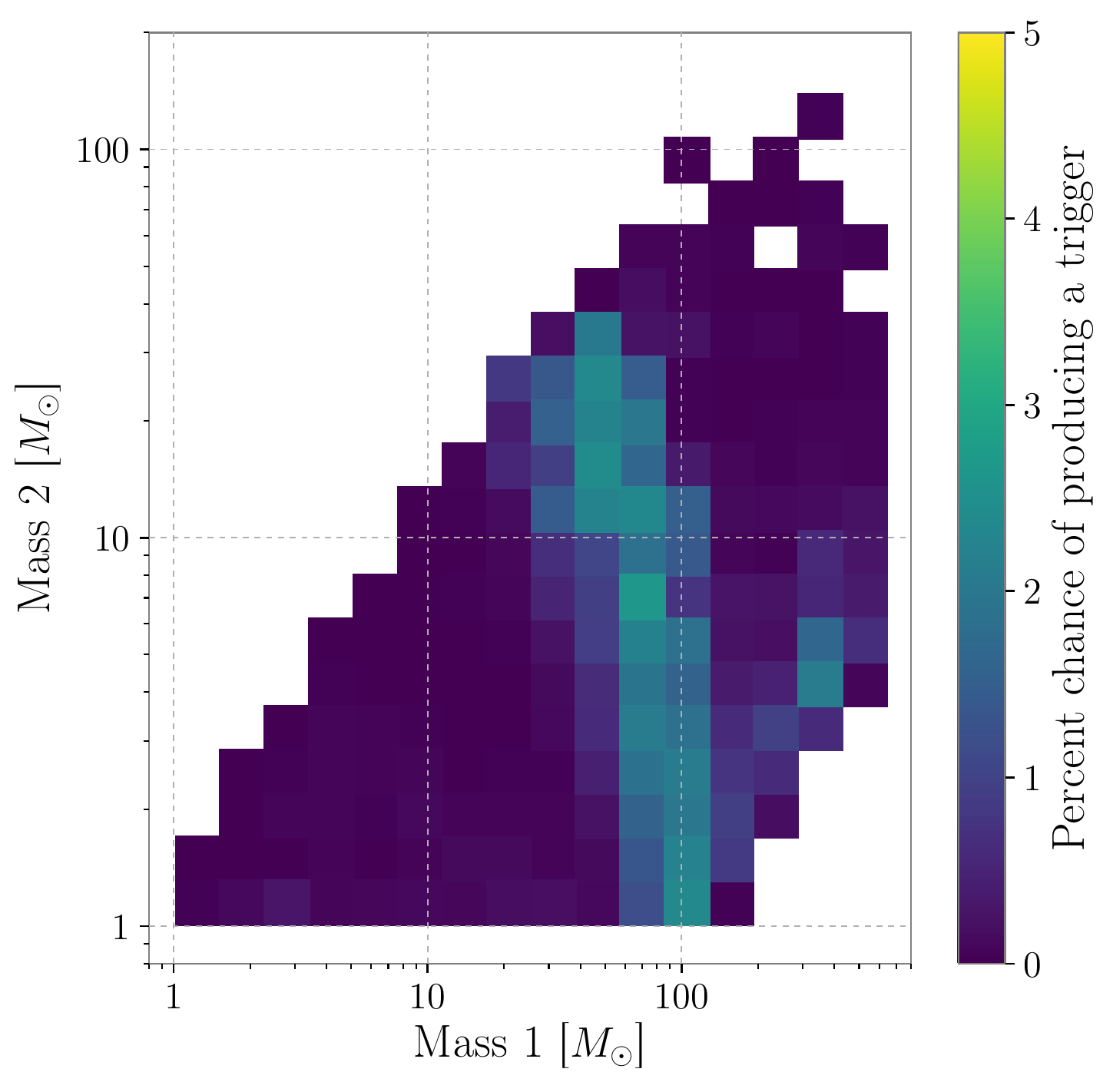}
        \caption{Blips}\label{fig:prob_MM_a}
    \end{subfigure} %
    \begin{subfigure}{.45\textwidth}
        \centering
        \includegraphics[width=\linewidth]{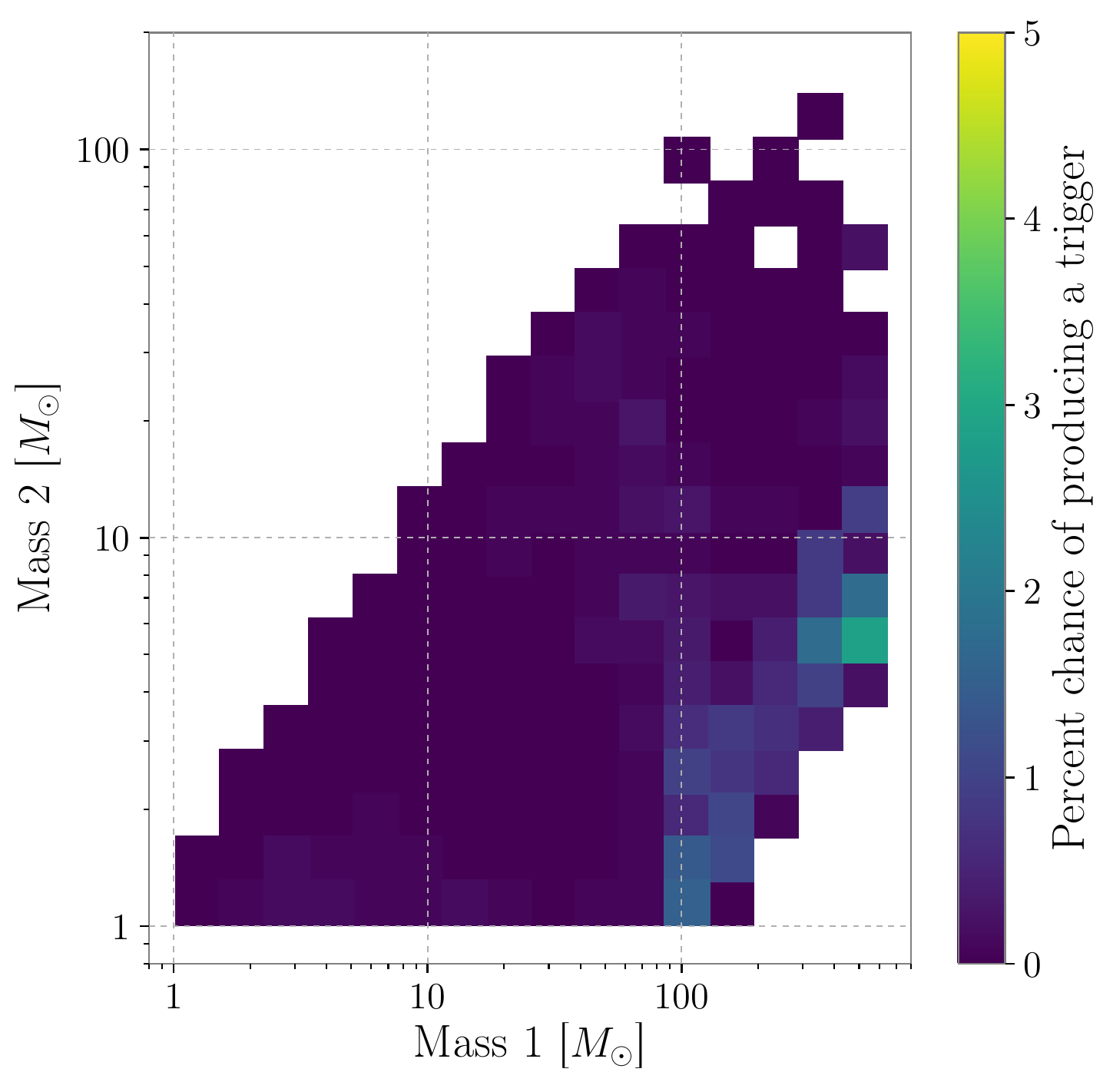}
        \caption{Koi Fish}\label{fig:prob_MM_d}
    \end{subfigure}
    \begin{subfigure}{.45\textwidth}
        \centering
        \includegraphics[width=\linewidth]{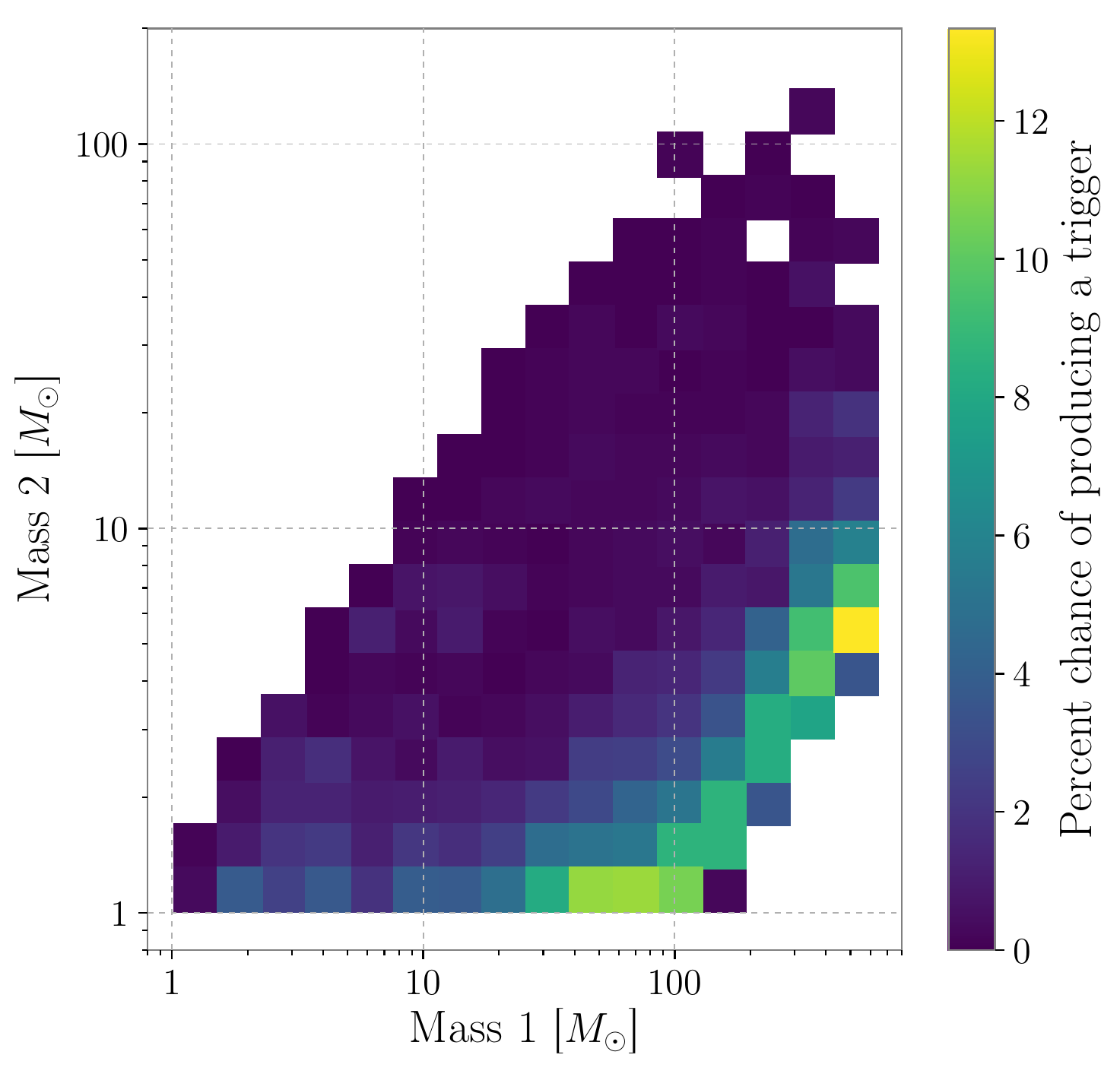}
        \caption{Scattered Light}\label{fig:prob_MM_b}
    \end{subfigure} %
    \begin{subfigure}{.45\textwidth}
        \centering
        \includegraphics[width=\linewidth]{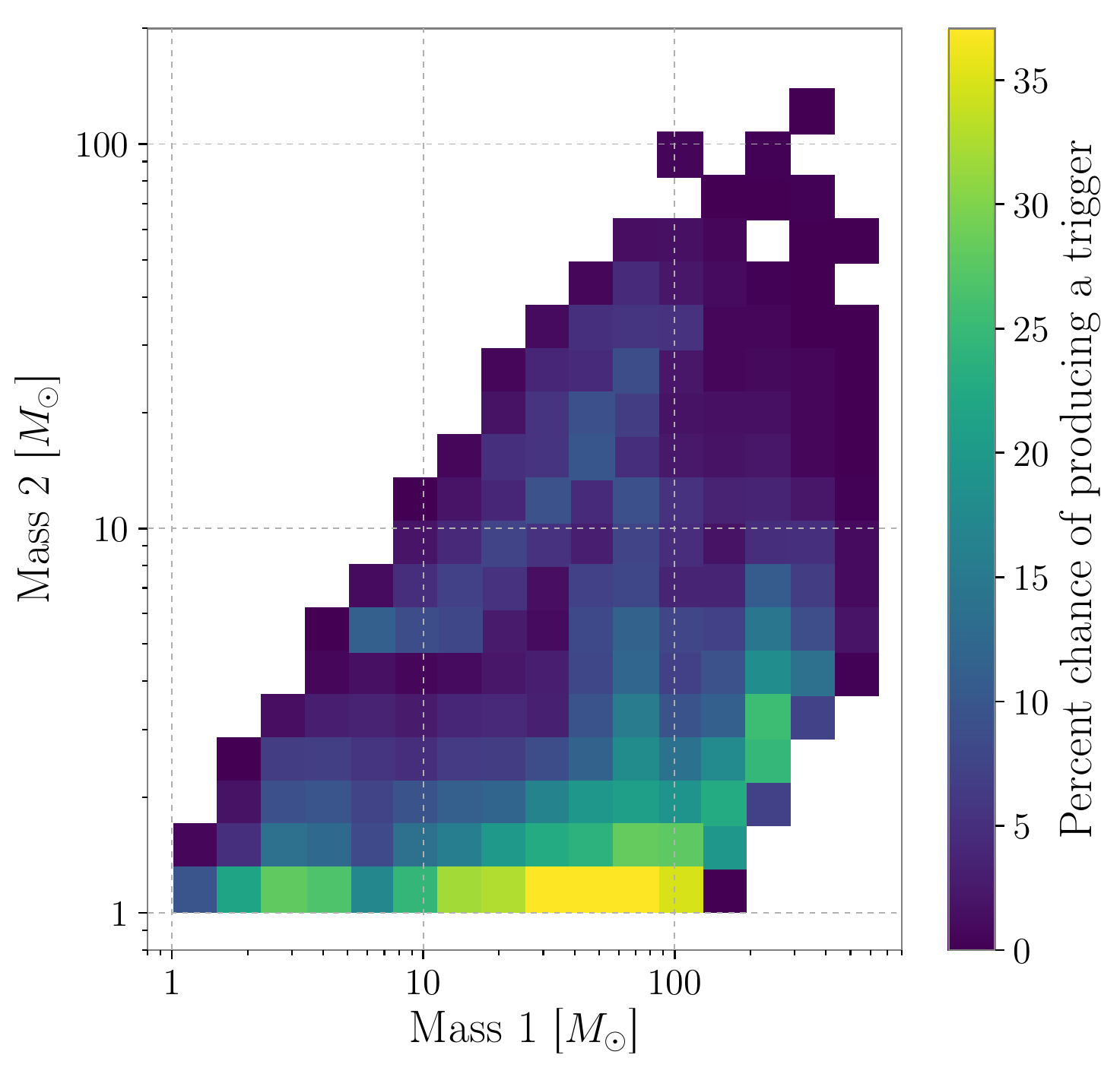}
        \caption{Scratchy}\label{fig:prob_MM_c}
    \end{subfigure}
\caption{Probability of producing a trigger above sine-Gaussian SNR of 7.0 in specific regions of the template bank parameter space for each glitch class.
Triggers are binned by component masses of the corresponding template.
Blips (top-left), koi fish (top-right),
scattered light (bottom-left) and scratchy (bottom-right) each show maximum probabilities
in different parts of the parameter space.
Note the differing range of values for each glitch class.
} 
\label{fig:prob_MM}
\end{figure}

In each template bank bin, we calculate the fraction of glitches in a given class
that produced at least one trigger above the chosen threshold that 
have parameters consistent with the bin in question.
Since each bin probability is calculated independently, the
sum over all bins is not bounded above by $1$.
In fact, it is possible for a given glitch instance to create triggers
recorded in multiple bins.
This allows the full extent of the glitch overlap with the template bank to be recorded. 
The results of this study can be seen in Figures 
\ref{fig:prob_MX} and \ref{fig:prob_MM}
for the four 
glitch classes considered in this work and for all of the template bank bins.
Figure \ref{fig:prob_MX} has templates binned by total mass ($M_{tot}$) 
and effective spin ($\chi_{eff}$)
while Figure \ref{fig:prob_MM} has templates binned by component mass.

Blip glitches show a strong clustering at the highest total mass in the template bank and 
maximally anti-aligned effective spins. 
In terms of component masses, the highest probability 
regions correspond to a primary mass of approximately 100 $M\odot$
and a smaller secondary mass. 
Notably, the highest probability for any region of the template bank is
below $5\%$. 
It is important to note that blips are one of the most common
classes of instrumental transient found in the detectors, 
so the low probability shown here may still impact the
sensitivity of the search. 
The rate of blip glitches is 1-2 per hour \cite{gwtc-1}, 
meaning that this rate corresponds to almost one significant 
blip-related trigger each day.

Triggers from koi fish glitches are most likely to be found
with high component masses, and maximal spins, both aligned and anti-aligned. 
Similar to blip glitches, these triggers correspond to 
short duration templates. 
However, koi fish triggers do not show the high probability
cluster for maximally anti-aligned templates
found for blip triggers.

The highest probability region for scattered light glitches 
is short duration highly aligned spin templates.
These templates experience a `hang-up' effect \cite{Healy:2018hu},
resulting in a template that can match the arch-like morphology
of a typical scattered light glitch.
Note that there is a large swath of parameter
space that has a noticeable response to a typical scattered light glitch.
This is likely a result of the large number of sources of scattered
light in the interferometer, which may
produce differing glitch morphologies.

Triggers coincident with scratchy glitches 
overlap well with a large range of templates,
with the highest probability clusters for maximally aligned and anti-aligned spins 
and total mass above $20 M_\odot$.
Projecting this result onto the component mass 
parameter space shows that this 
region also corresponds to
templates with high mass ratios, such as those from 
neutron star - black hole (NSBH) systems. 
In addition to high mass triggers, the component mass parameter space
also shows a high probability of producing a trigger with component masses
of $3 M_\odot$ and $1 M_\odot$.
While scratchy glitches are rarer than the other classes 
identified in this work, the likelihood of a single glitch
producing a significant trigger is high for a wide range of parameters,
a contrasting situation to blips and koi fish.
This is partially due to the typical timescale of a scratchy glitch, 
lasting up to multiple minutes, 
as opposed to 
Blip and koi fish glitches that last tenths of seconds.
While this does increase the chance of a trigger being due to chance, 
the excess is significantly higher than
we would expect from colored Gaussian noise alone.
Scratchy glitches are also the only class of the four
surveyed in this work that are shown to have a high probability
of producing a significant trigger with a total mass below $20 M_\odot$.

These results on their own can not preclude the
possibility of a genuine signal to occur in coincidence
with a glitch (as was the case for GW170817).
However, for each glitch class, there does appear to be a range of parameters
for which a candidate trigger in time coincidence would
likely be caused by the presence of the glitch. 
This clustering allows these results to be used to quickly follow up
candidates after they are identified by the search to 
understand if the recovered parameters are likely to be produced 
by the glitch in question. 

For example, if a significant gravitational-wave candidate
was identified during a time period classified by Gravity Spy as
a blip, the candidate trigger is unlikely to be related 
to the observed instrumental artifact unless trigger is
a high mass, anti-aligned template.
Conversely, a trigger candidate coincident with a glitch that 
has parameters consistent with a high probability region
of the glitch would warrant
additional investigations to understand a possible connection between
the artifact
and candidate trigger

Notably, the mass and spin parameters of previously
observed BBH and BNS signals \cite{gwtc-1} are not consistent with any
of the highest probability regions for any of the four glitch
classes examined in this work.
This suggests that it unlikely
for any of these common glitch classes
to mimic a signal from the currently observed 
gravitational-wave population.

\section{Utilizing glitch classification in significance estimates}\label{sec:contam} 

As discussed in the previous section, 
the known correlations between specific template parameters and 
glitch classes do not preclude the possibility of a real astrophysical signal
occurring in coincidence with a glitch. 
In this scenario, it may be possible to include 
the additional information we have about the expected
overlap with the candidate trigger parameters
and the glitch population to re-evaluate the significance
of the candidate.
In this section we outline a procedure
to calculate the significance
of a trigger found in time coincidence
with a time classified by Gravity Spy as belonging
to a specific glitch class. 
Rather than producing a ``yes or no'' result on whether a 
gravitational-wave candidate is ``caused'' by the presence
of a glitch, we can re-estimate the significance of a candidate
based on the expected rate of triggers due to the glitch.
This approach decreases the risk that a astrophysical signal
will be discarded due to data quality issues.
We focus on blip glitches as a test case since 
they are one of the most common 
glitches in both detectors and have very defined regions
of the parameter space where overlap between signals and glitches
occur.

\begin{figure}[t]
\centering
    \includegraphics[width=.45\textwidth]{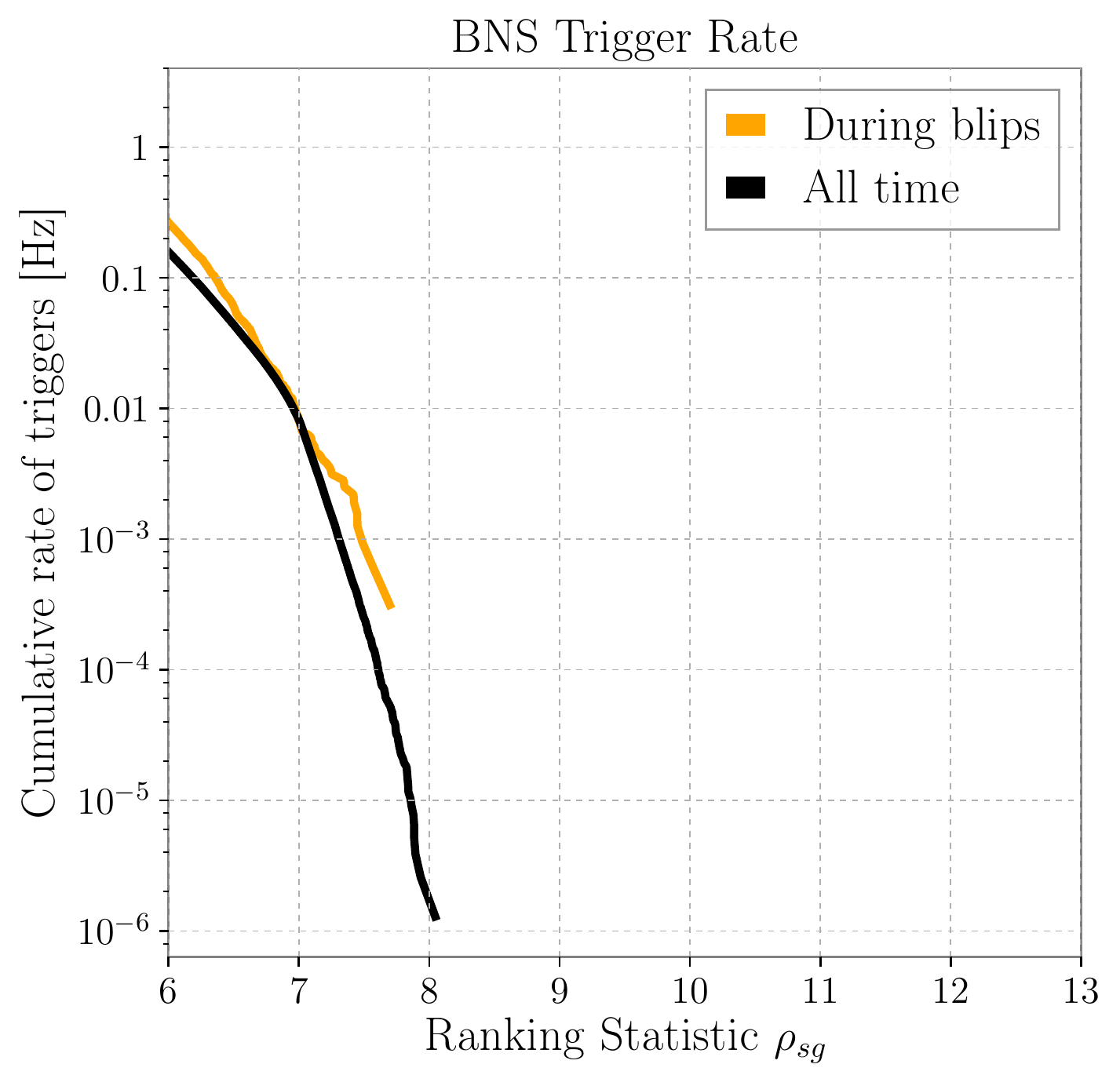}
    \includegraphics[width=.45\textwidth]{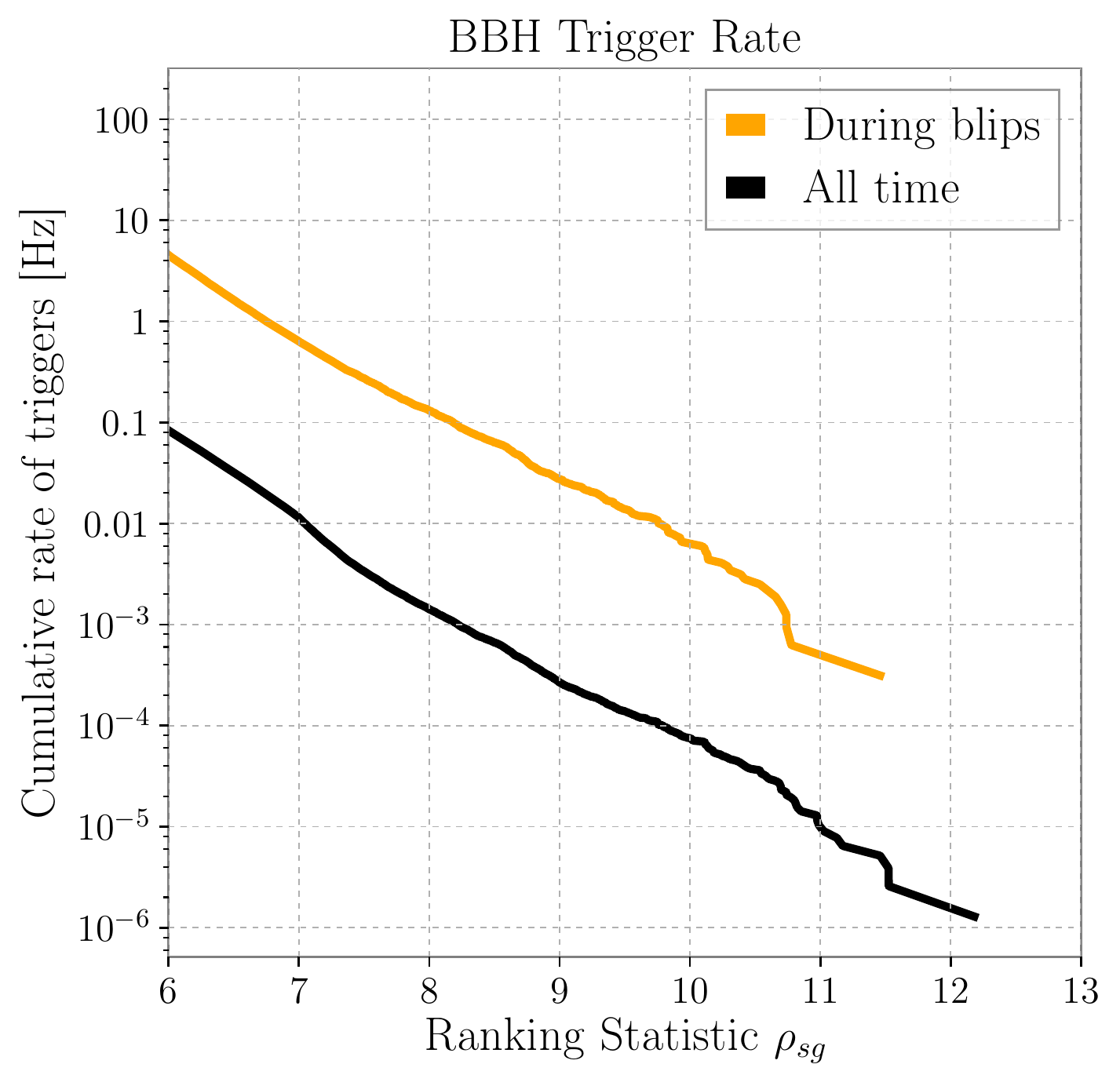}
    \caption{A comparison of the false alarm rate (FAR) based on all times %
    in the analysis and only times coincident within 2 seconds of blip. %
    The trigger rate is much higher during blip periods for BBH triggers, %
    supporting the previous conclusion that there is an increased chance %
    of producing a trigger coincident with this source of noise. %
    Left: Difference in trigger rate for BBH triggers. %
    Right: Difference in trigger rate for BNS triggers.}
    \label{fig:back}
\end{figure}

To first demonstrate how glitching changes the trigger rate
in different parts of the parameter space, we compare
trigger rates from the entire analysis period versus only times
that are known to contain blip glitches. 
We then sort triggers from each period into a long duration (referred to as BNS) category
($M_{chirp} < 2.0 M_\odot$) 
and a short duration (BBH) category ($M_{chirp} > 5.0 M_\odot$).
A comparison of the rate of triggers versus network ranking statistic 
$\tilde{\rho}_{sg,net}$ for the entire analysis period and
around blip glitches is shown in Figure \ref{fig:back}.
If we examine the trigger rate during the entire analysis
period versus during time periods within 2 second of blip glitches, 
we see that there is indeed an increased rate of triggers at fixed ranking statistic
for times around blip glitches.
Furthermore, this increase is only apparent for the BBH category, 
with only minimal increases for the BNS category.
This agrees with our expectation from the previous section that blip glitches
are correlated with high mass triggers.

As was the case with GW170817, it may still be possible to 
detect a gravitational-wave signal during a time period
corrupted by glitching. 
A critical component of this detection was establishing
that the observed glitch could not have accounted for the
BNS signal in the data \cite{GW170817,Pankow:2018mit}. 
In order to facilitate significance estimates of
additional candidates, Gravity Spy classifications,
combined with our knowledge of the overlaps
between template parameters, can be used
to evaluate if the candidate trigger
is significant despite the association with 
a known class of glitch.

Blip glitches present a clear use case for this
follow up.
This class is known to impact only 
an isolated part of the parameter space.
Specifically, we would expect that low mass BNS triggers
would be uncorrelated, while high mass BBH triggers
may be correlated to the presence of a glitch.
In order to account for the expected variation in the 
background distribution across the template bank, 
the PyCBC pipeline includes parameter dependent background reweighing
that measures the rate of triggers with respect to template duration
and downranks templates that are shown to occur
more frequently \cite{Nitz:2017stat}.
As BNS and BBH signals have vastly different
template durations, we would expect both classes of signals
to be affected differently by the inclusion of this term
to the ranking statistic when compared against our expected
triggers from blips. 

\begin{figure}[t]
\centering
    \begin{subfigure}{0.45\textwidth}
        \centering
        \includegraphics[width=\linewidth]{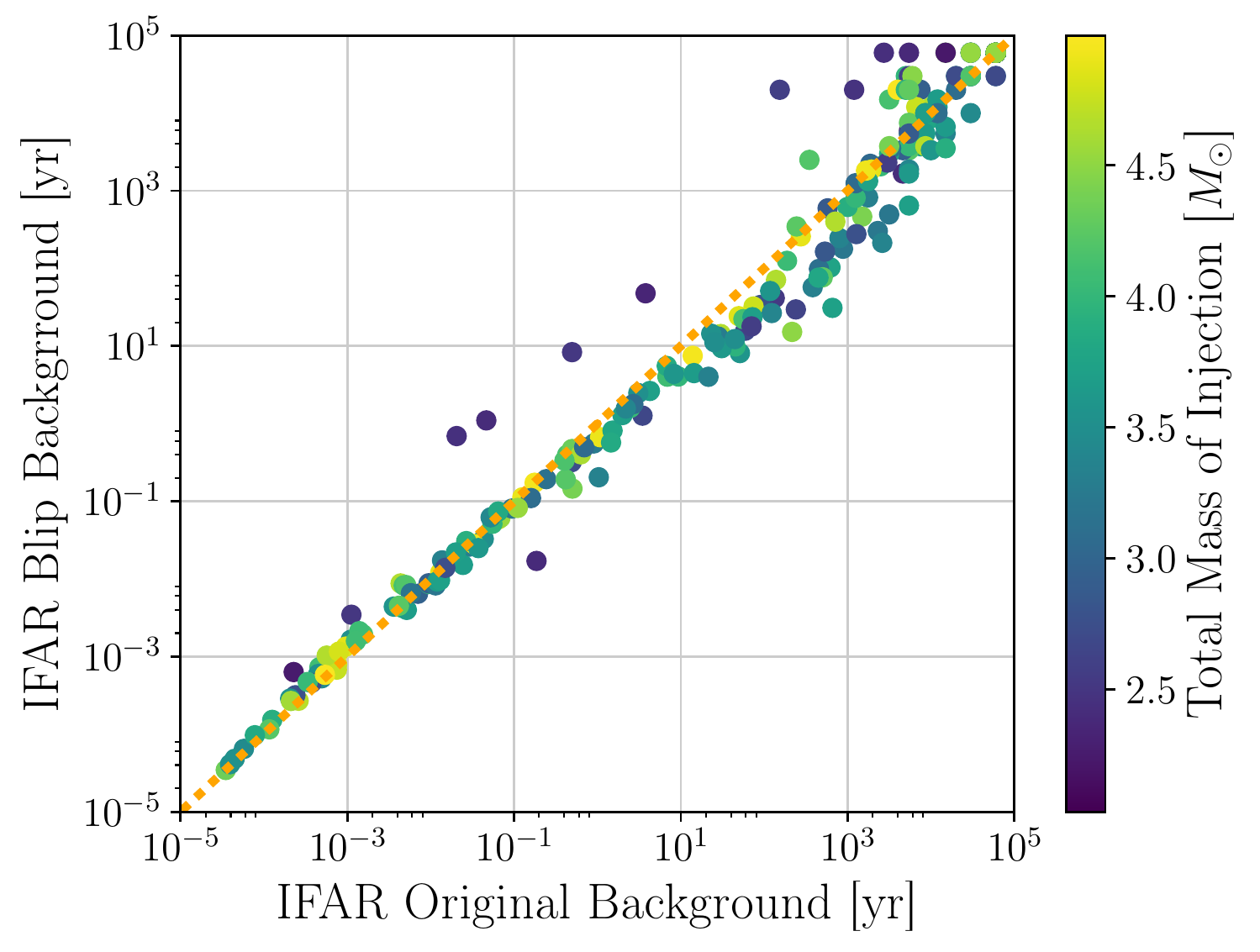}
        \caption{BNS Injections}\label{fig:ifar_bns}
    \end{subfigure} %
    \begin{subfigure}{.45\textwidth}
        \centering
        \includegraphics[width=\linewidth]{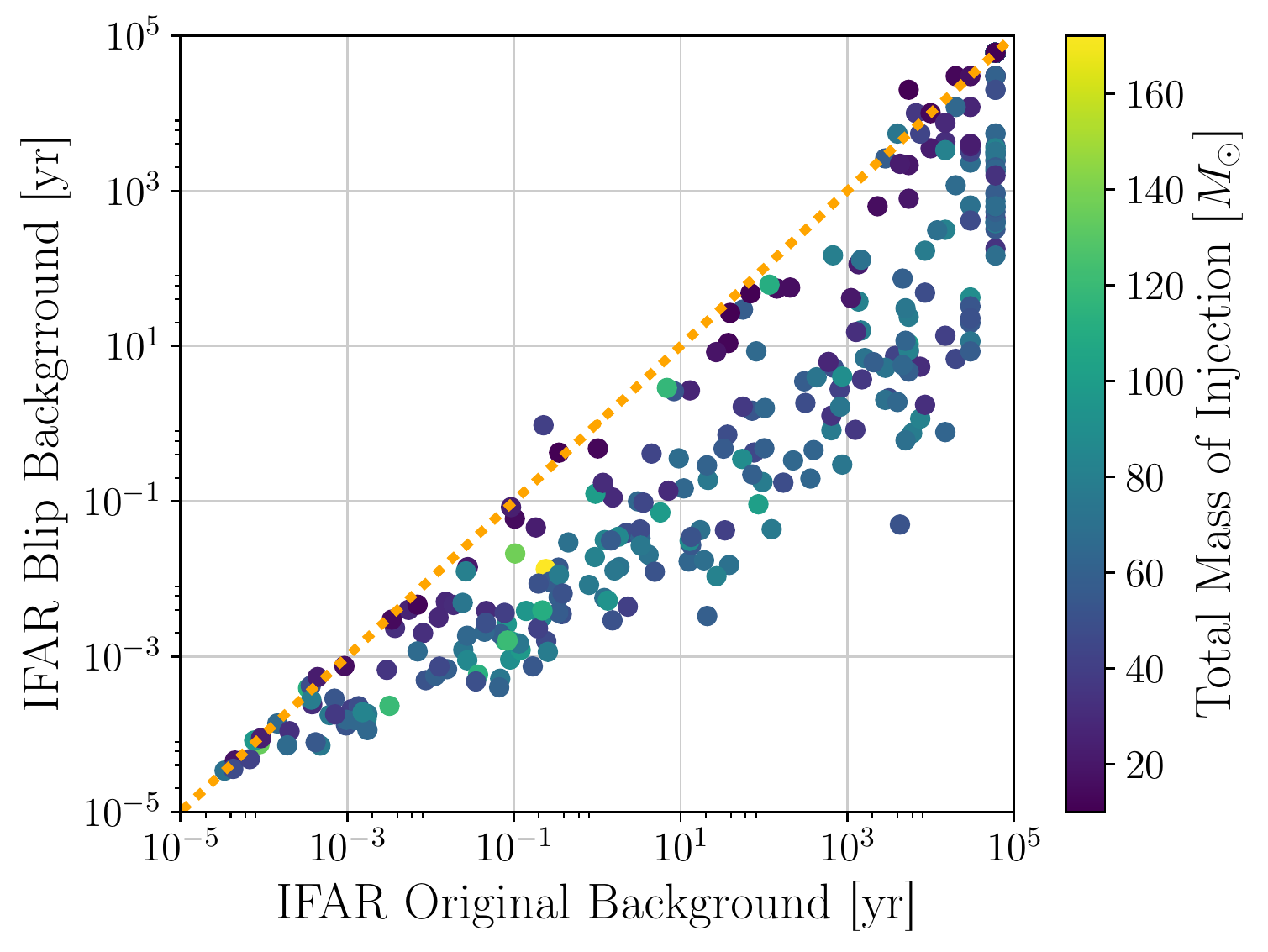}
        \caption{BBH Injections}\label{fig:ifar_bbh}
    \end{subfigure}
    \caption{Recovered IFAR for a set of injections compared against the %
    background from all time during an analysis (x-axis) versus the %
    background during only blip times (y-axis). In each plot a 1-1 line %
    is included for reference. %
    For BNS injections (left), there is no effect on the recovered IFAR. %
    For BBH injections (right), the high rate of triggers with similar %
    template duration during blip times reduces the significance of the %
    injections. %
    This provides a natural way to evaluate a candidate trigger %
    that is coincident with a known source of noise in the detector.}
\label{fig:ifar}
\end{figure}

To demonstrate how the significance of each
signal model is affected by a correlation with blip glitches,
we perform a series of astrophysical 
software injections of both BNS and BBH signals
into aLIGO data that are similar to the population 
of BNS and BBH signals previously detected by aLIGO and aVirgo. 
We first calculate the inverse false alarm rate (IFAR)
of each injection using the background distribution measured
from a 5 day analysis period
as a control.
We empirically measure the distribution of single 
detector background triggers during times period flagged
by Gravity Spy as blips with confidence $>0.95$, 
and use this as the input for the background re-weighing
procedure. 
We then reevaluate the IFAR of each injection with this
new background distribution.
A comparison of the recovered IFAR for each injection
before and after including this correction based on the expected background
distribution of blips can be seen in Figure \ref{fig:ifar},
along with a 1-1 line indicating where the recovered
IFAR is consistent between the two cases.

Comparing the two injection sets, there is a clear difference 
in the distribution with respect to the 1-1 line. 
BNS injections were recovered at approximately the same IFAR in
both cases, showing that times corrupted by blips
are equally likely to produce a BNS trigger as an 
average time. 
The BBH injections, on the other hand, have a clear separation 
from the 1-1 line. 
Specifically, the IFAR of injections 
recovered is lower when a background based on blips
is used versus the current search configuration. 
Since blip times are more likely to produce triggers with
BBH parameters, these injections are naturally downranked. 
However, sufficiently loud injections are still recovered at IFAR of greater
than 1 year, which would be sufficiently significant
to separate itself from the background. 
This allows signals which are inconsistent with the
expected glitch behavior
to be identified as a candidate of interest.

While only blips were studied in this section, 
this procedure can be repeated for all glitch classes
that are sufficiently common for an expected
background distribution to be measured. 
This method can provide further quantitative evidence that
a candidate trigger is unlikely 
to be related to the instrumental artifact
that it happens to occur in time coincidence with. 
In general, the presence of a glitch
does not preclude the possibility that a candidate trigger
is astrophysical, but only reduces
the likelihood of astrophysical origin as compared to 
a candidate trigger that does not overlap a known 
instrumental artifact. 
As a high number of significant triggers
have occurred during periods of transient noise \cite{gwtc-1},
the methods outlined in this work will be able to 
be applied to numerous gravitational-wave events.

In addition to being used in validation procedures of significant
candidates, 
this method can be implemented in the ranking statistic
internal to the search pipeline.
One possible technical solution is to separately
evaluate triggers during times categorized by Gravity Spy as 
glitches and times where no glitch is identified.
This would allow the search to benefit from an increased
sensitivity during time periods where no glitching occurs, and
to more accurately rank candidates related to glitching.
Down ranking triggers during glitches instead of removing them
from the analysis has the benefit, as
compared to current data quality veto methods, of not precluding the 
possibility of detecting a signal during a glitch.
For glitches without known witnesses, 
such as blips, 
this method also allows for Gravity Spy classifications
to be incorporated into significance estimates
of gravitational-wave candidates.
If an astrophysical signal was  
classified as a glitch due to a highly similar
morphology with a glitch class, 
this method would not prevent detection. 
Further studies evaluating the safety of this method, 
and the utilization of Gravity Spy classifications
to rank candidate triggers, are warranted.

One limitation of this method is that it relies
upon the specific parameters used to model
the background distribution of the matched filter search. 
As has been shown in this work, 
classifying PyCBC triggers with
one single variable will not be able to fully
differentiate triggers due to glitches
and triggers due to genuine gravitational waves. 
This method will likely be more effective if
additional parameters are used
(such as the effective spin and total mass).
Detailed investigations into what parameters
would be best for each glitch class may be 
resolved in future investigations. 

\section{Discussion} 

This work emphasizes how similarities 
between common glitches and astrophysical signals in aLIGO data
can present challenges in validating 
gravitational-wave detections.
This is especially problematic for novel sources including 
mergers of intermediate mass black holes 
\cite{O1_IMBH,IMBHB:S5} 
and neutron star - black hole \cite{O1:BNS_NSBH} systems. 
As each of these regions of the current template bank are impacted differently by 
each of the glitch classes, 
there is unlikely to be a single method to efficiently differentiate these novel sources
from common instrumental artifacts. 
Focused work to design consistency tests that account for 
known problematic glitch morphologies is needed.
Alternatively, developing robust mitigation techniques for each
of these common glitch classes will have tangible effects on the overall
sensitivity of the searches. 

At the present, the quantifiable metrics developed in this work can also be used to guide 
event validation of candidate triggers. 
When evaluating whether to initiate a search for an electromagnetic
counterpart, being able to predict
whether a significant trigger is likely due to the presence of a 
common glitch will allow more informed and prompt alert updates. 
As many EM counterparts to a gravitational wave signal 
occur within minutes of merger \cite{Cannon2011Early}, 
quick follow up is critical. 
Gravity Spy classifications are currently utilized
in automated follow up in LIGO-Virgo's third observing run 
and the results of this study can be used to translate these classifications
into easily used metrics to determine the likelihood of the candidate being
related to a common glitch. 
Since the regions where a glitch is most likely to produce a trigger
correspond with regions where a lower event rate is expected, 
understanding if a candidate of interest is a rare astrophysical signal
or a common glitch is especially important to guide 
astronomical observations. 

Once aLIGO reaches design sensitivity and the rate of detections
increases, signals found near noise transients in the data will 
become a much more common situation. 
Already in the recent results from
O2, a significant fraction of marginal triggers
have been identified in time coincidence
with noise transients \cite{gwtc-1}.
This trend has continued during the third observing run,
where a number of open alert candidates
have been retracted due to data quality concerns 
\cite{GCN24591,GCN24655,GCN25301}.
Future progress in addressing these issues
will be facilitated both by continual 
instrumental work to the reduce the rate of glitches
as well as by further studies of how to distinguish 
glitches from genuine signals.

\section{Acknowledgments}
We would like to thank the Gravity Spy team and the
Detector Characterization group for 
identifying and characterizing the noise sources addressed in this work,
as well as the PyCBC search group for development of the
pipeline. 
We also thank Scott Coughlin, Joshua Smith, and Thomas Massinger
for helpful discussions. 
Computing support for this project was provided by the LDAS computing cluster 
at the California Institute of Technology.

DD, LW, and PRS acknowledge support from NSF award PHY-1607169.
LIGO was constructed by the California Institute of Technology and Massachusetts 
Institute of Technology with funding from the National Science Foundation, and 
operates under cooperative agreement PHY-0757058. 
This work carries LIGO Document number P-1900372.

\end{document}